\documentclass[prl,aps,twocolumn]{revtex4-2}
\usepackage{amssymb}
\usepackage{graphicx}
\usepackage{placeins}
\usepackage{amsmath}
\usepackage{color} 

\definecolor{darkgreen}{rgb}{0,0.7,0}  

\usepackage[normalem]{ulem} 
\usepackage{xr-hyper}
\usepackage{microtype}
\usepackage[table]{xcolor}
\definecolor{lightgray}{gray}{0.9}
\usepackage{booktabs}

\usepackage[caption=false]{subfig}
\usepackage{hyperref}

\begin{document}
\title{Dispersive Shock Waves in a 1D Quantum Liquid}
\author{Philipp Sch{\"u}ttelkopf\textsuperscript{1}}\email{philipp.schuettelkopf@tuwien.ac.at}
\author{Mohammadamin Tajik\textsuperscript{1,3,4}}
\author{Federica Cataldini\textsuperscript{1}}
\author{Si-Cong Ji\textsuperscript{1}}
\author{Igor Mazets\textsuperscript{1}}
\author{Sebastian Erne\textsuperscript{1}}
\author{Nataliia Bazhan\textsuperscript{1}}
\author{Mojtaba Alyannezhadi\textsuperscript{1}}
\author{Mostafa Alyannezhadi\textsuperscript{1}}
\author{J\"org Schmiedmayer\textsuperscript{1}}
\author{Frederik M{\o}ller\textsuperscript{1,2}}
\affiliation{
\textsuperscript{1}Vienna Center for Quantum Science and Technology (VCQ), Atominstitut, TU Wien, Vienna, Austria \\
\textsuperscript{2} Institute of Science and Technology Austria, Am Campus 1, 3400 Klosterneuburg, Austria \\
\textsuperscript{3} Division of Biology and Biological Engineering, California Institute of Technology, Pasadena, CA 91125 \\
\textsuperscript{4} Max Planck Institute of Molecular Cell Biology and Genetics, 01307 Dresden, Germany
}
\date{\today}

\begin{abstract}
We implement a moving boundary condition in a 1D quantum liquid to study nonlinear wave breaking and its regularization by dispersion. Programmable optical potentials allow us to compress a weakly interacting ultra-cold Bose gas trapped on an atomchip at tunable speeds of up to three times the speed of sound and subsequently measure the quasi-in-situ density distribution to extract the shock wave edge dynamics. We resolve both leading and trailing edge velocities and observe a shock wave width that increases linearly in time, which are distinguishing features of dispersive shock waves, consistent with asymptotic predictions using Whitham's method. Quantitative agreement is found with finite temperature non-polynomial Schrödinger equation simulations, taking into account the imaging process and the finite height of the piston potential. Our results constitute a controlled, quantitative test of dispersive shock dynamics in a 1D quantum fluid and demonstrate that the coarse-grained dispersive-shock phenomenology remains robust even as the microscopic dynamics depart from the strictly integrable 1D regime.
\end{abstract} 

\maketitle

\textit{Introduction $-$} In nonlinear media steepening of wave fronts leads to a so-called \textit{gradient catastrophe}, requiring a mechanism to regularize the ensuing wave breaking. Classical fluids exhibit viscosity, resulting in the formation of viscous shocks and entropy production, whereas in dissipationless media dispersion yields an expanding rank-ordered train of oscillations, namely a dispersive shock wave (DSW) \cite{nonlinear_DSW_whitham_1965, sw_disp_hydro_gurevich_1984, DSW_discontinuity_whitham_gurevich1987dissipationless, linear_and_nonlinear_waves_whitham2011_BOOK, dispersive_shock_waves_mod_theory_el_hoefer_2016, Piston_DSW_analytics_Hoefer_2008}. 

Ultracold dilute Bose gases present a powerful platform for investigating quantum many-body dynamics \cite{Ultra-cold-gases_Bloch_2008,Review_one_dim_bosons_Rigol_2011,quantum_simulators_ultracold_gases_Bloch2012,Trapped_1d_gases_Petrov_2000}. Confined to 1D, they become close to integrable \cite{hydrodyn_of_integr_systems_Bulchandani_2017} and effectively dissipationless over experimental timescales. This makes them an ideal testbed for dispersive phenomena, while offering precise experimental control. Following recent progress in the dynamics of 1D systems, investigation of higher-order hydrodynamic effects has attracted a lot of interest \cite{ emergent_hydro_integr_sys_in_q_sys_out_of_equi_Castro_Alvaredo_2016,GHD_transport_xxz_chains_Bertini_2016,hydrodyn_diffusion_integrable_sys_Nardis_2018,correspondence_Whitham_GHD_GPE_Bettelheim_2020,finite_T_q_SW_Kheruntsyan_2020,GHD_1d_Bouchoule_2022,one_dim_DSW_study_varying_interaction_density_Kheruntsyan_2023,doyon2023generalized, whitham_mller_2024,diffusive_lenght_scales_1d_bose_gas_moller_2024,q_fluc_1_dim_SW_Jacopo_2025,Drude_weights_schuettelkopf_2026}. The ensuing rich nonequilibrium dynamics at the wave breaking offer insights beyond the ballistic (Euler) regime. To that end, the canonical piston protocol constitutes a direct, controlled, and quantitative test of dispersive shock dynamics.

Dispersive shock waves have been reported in nonlinear optics for an initial discontinuity \cite{ optical_intensity_shock_light_pulse_fiber_NLS_rothenberg_1989, shock_gaussian_laser_pulse_Stefano_2007,DSW_Bipartition_light_pulse_Stefano_2017} and during four-wave mixing \cite{undular_bores_4_wave_mixing_Trillo_2014}. 
However, in systems of ultracold Bose gases, investigations of shock dynamics have been limited to 3D \cite{EXP_3d_BEC_vortices_Dutton_2001, EXP_sound_waves_rotating_3DBEC_Simula_2005, DSW_3D_Exp_Hoefer_2006}, albeit some highly anisotropic configurations \cite{observation_sw_BEC_meppelink_2009, DSW_formation_merging_splitting_BEC_chang_2008, Dissipative_shock_wave_Mossman2018}. Thus, the ensuing dynamics exhibited dissipation in the form of formation of topological defects. While previous works present dynamics consistent with DSWs \cite{DSW_3D_Exp_Hoefer_2006, observation_sw_BEC_meppelink_2009, DSW_formation_merging_splitting_BEC_chang_2008}, they do not resolve both edge velocities, the shock width growth or the behavior at the critical velocity. 

Here we implement the canonical piston problem in a strongly confined quasi-1D quantum liquid using a programmable optical wall and probe the shock dynamics for a wide range of piston speeds. We analyze coarse-grained density profiles after time-of-flight (TOF) to directly measure both shock edge velocities and the shock region width, a distinguishing feature of DSWs \cite{nonlinear_DSW_whitham_1965, kamchatnov2000nonlinear,linear_and_nonlinear_waves_whitham2011_BOOK, dispersive_shock_waves_mod_theory_el_hoefer_2016, Kamchatnov_2021}. We subject our effectively dissipationless system to large perturbations, pushing local densities beyond the strict 1D-condition \cite{Trapped_1d_gases_Petrov_2000, Review_one_dim_bosons_Rigol_2011, Salasnich_2002_eff_wave_eq}, testing dispersive nonlinear wave dynamics. Further, our work constitutes a quantitative test of Whitham's method \cite{nonlinear_DSW_whitham_1965, Whitham_mod_theory_orig_1_1965, dispersive_shock_waves_mod_theory_el_hoefer_2016, linear_and_nonlinear_waves_whitham2011_BOOK}, which provides an asymptotic solution to the dispersive shock wave in the piston problem. \\

\begin{figure*}[!t]
    \centering
    \includegraphics[width = 17.2 cm]{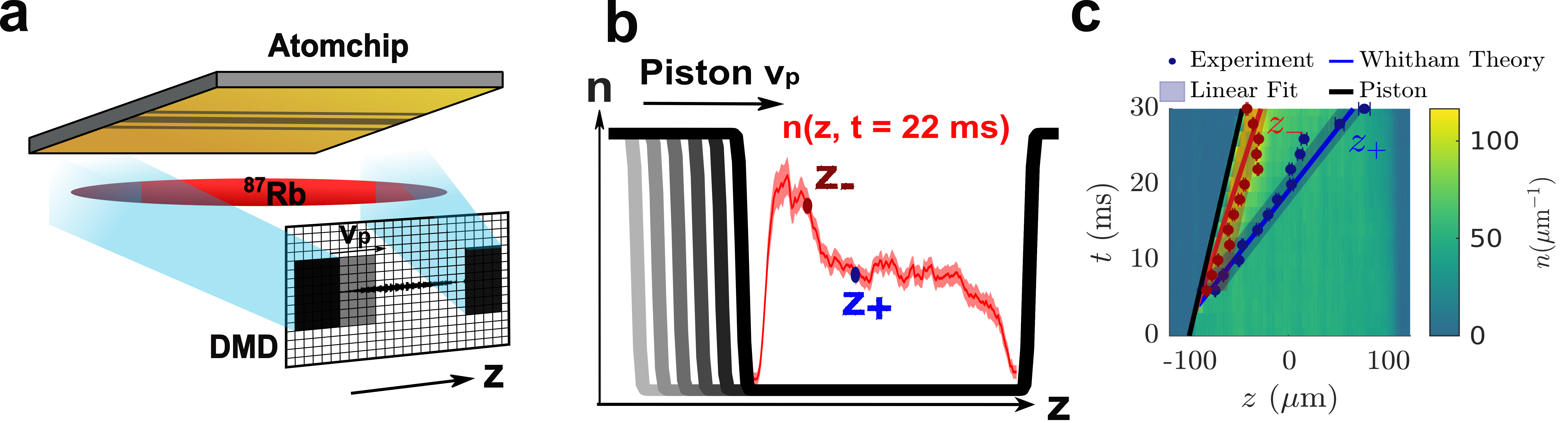}
    \caption{\label{fig:exp_piston_problem}
    \textbf{Illustration of the experimental protocol and measured density carpet.}
    \textbf{(a):} Ultracold $^{87}$Rb atoms are trapped in a 1D confinement using an atomchip. Superimposing (blue detuned) optical dipole potentials created by a DMD enables the creation of arbitrary dynamic longitudinal potentials, including 
    \textbf{(b):} the realization of a moving piston by successively turning on columns of pixels. The averaged density profile $n(z,t)$ is shown in red (shaded: standard error), with extracted shock edge positions $z_\pm$.
    \textbf{(c):} Density carpet measured during a piston protocol: $n(z,t)$ as a function of position and time. A piston with constant velocity $v_\mathrm{p}$ compresses the condensate from the left. The extracted shock edge positions $z_\pm$ are shown as red dots and the corresponding linear fits $v_\pm$ are shown as solid blue/red lines with confidence intervals as shaded regions. Theoretical predictions using Whitham's method are shown as solid lines. Measurement time is limited by the finite system size.
    }
\end{figure*}

\textit{Experimental Setup $-$ }Our experimental system is a one-dimensional ultracold Bose gas of $4-10 \times 10^3$ $^{87}$Rb atoms trapped below an Atom Chip~\cite{reichel2011atom}. The chip produces a cigar-shaped magnetic trap with $\omega_\perp = 2 \pi \times 1.38~\mathrm{kHz}$ along the two transverse directions, satisfying the 1D condition $ \hbar\omega_{\perp}>k_BT , \mu $, where $\mu$ is the chemical potential, $k_B$ Boltzmann's constant, and $T$ the temperature, which is typically in the range $20-90$ nK. The interaction strength is quantified by the dimensionless Lieb-Liniger parameter $\gamma = m g_{1\mathrm{D}} / (\hbar^2 n_{\text{1D}})$~\cite{lieb1963exact}. In our experiment, $\gamma \approx 0.002$, which, combined with the low temperature, places the gas deeply in the quasi-condensate regime. To implement arbitrary dynamic longitudinal potentials, we utilize a Digital Micromirror Device (DMD) combined with a blue detuned laser ($\lambda = 766.7~\mathrm{nm}$) to project pre-engineered intensity patterns onto the gas along the weak longitudinal confinement in $z$-direction~\cite{Tajik:19}, see Fig.~\ref{fig:exp_piston_problem}(a).\\
\FloatBarrier
\textit{The piston $-$ }To implement the piston protocol we prepare a 1D thermal state directly into a box potential. At $t=0$ we start to successively turn on columns of DMD pixels next to the wall, which locally increases the light intensity, moves the wall, and pushes the atoms inward at a constant velocity, thus implementing the piston. In the resulting dynamics, atoms accumulate at the edge of the piston and a shock region forms, which is characterized by a rapidly oscillating density profile. In classical hydrodynamics, a balance of dissipation and nonlinearity leads to a viscous monotonic shock region of constant width \cite{linear_and_nonlinear_waves_whitham2011_BOOK}. In contrast, in the absence of dissipation, dispersion gradually fans out the oscillations depending on their amplitude, resulting in an oscillating growing shock region \cite{linear_and_nonlinear_waves_whitham2011_BOOK}. Towards the leading edge, short wavelength small-amplitude linear oscillations are predicted on the unperturbed background, while the trailing edge matches a (large-amplitude) soliton on the compressed region \cite{linear_and_nonlinear_waves_whitham2011_BOOK}. The experimental protocol is illustrated in Fig.~\ref{fig:exp_piston_problem}.\\

\textit{Measurement $-$} We measure the quasi-in-situ linear density profile $n(z,t)$ via absorption imaging after 2 ms TOF expansion, repeating the experiment for several evolution times $t$ and piston speeds $v_\mathrm{p}$ for 50-80 repetitions. We determine the leading and trailing edges of the shock region for intermediate evolution times $t$, that is, after the shock region has expanded to an observable size, but before the train of waves created by the piston has reached the opposite wall of the box. An example profile is shown in Fig.~\ref{fig:exp_piston_problem}(b). We identify an unperturbed region matching the initial density distribution far from the action of the piston. The leading edge $z_+$ of the shock region is found where a fitted error-function of the relative density $n(z,t)-n(z,0)$ exceeds the standard error of the density distribution (details see Supplementary Material \cite{Supplementary_Material}). The trailing edge $z_-$ is found at the edge of the error function, except close to the critical velocity $1.4 v_\mathrm{s}<v_\mathrm{p}<2.2v_\mathrm{s}$, where $z_-$ coincides with the position of the piston. Notably, we find that the wave fronts $z_\pm$ propagate with constant velocity, see Fig.~\ref{fig:exp_piston_problem}(c), enabling the straightforward extraction of the corresponding DSW velocities $v_\pm$ via a linear fit.\\

\textit{Framework $-$} The piston problem is typically studied using Whitham's method applied to the zero temperature Gross-Pitaevskii equation (GPE). This yields an asymptotic description of the coarse-grained shock wave envelope in the form of a traveling wave, from which the oscillations can be reconstructed (see Supplementary Material\cite{Supplementary_Material}) \cite{Piston_DSW_analytics_Hoefer_2008,dispersive_shock_waves_mod_theory_el_hoefer_2016, Kamchatnov_2021, nonstationary_KdV_gurevich_pitaevskii_1973, sw_disp_hydro_gurevich_1984, linear_and_nonlinear_waves_whitham2011_BOOK, nonlinear_DSW_whitham_1965}. Because the shock wave microstructure lies below our imaging resolution (healing length $\xi_\mathrm{h} \approx 0.26~\mu\mathrm{m}$, imaging point-spread-function width $\sigma_{\mathrm{PSF}}\approx 3~\mu\mathrm{m}$), the highly oscillatory wave train of the DSW cannot be imaged directly. Additionally, finite temperature effects and experimental fluctuations decrease the contrast of the DSW, especially towards the leading edge \cite{finite_T_q_SW_Kheruntsyan_2020, whitham_mller_2024}. Furthermore, free evolution during TOF expansion mixes phase and density (see Supplementary Material \cite{Supplementary_Material}): large phase gradients at the trailing edge of the DSW lead to an increase in observed atomic density \cite{whitham_mller_2024}, whereas the leading edge is minimally affected, as phase gradients are small. These experimental constraints and factors are not included in Whitham's treatment of the piston. Therefore, we compare experimental data to finite-temperature simulations of the quasi-1D non-polynomial Schrödinger equation (NPSE) \cite{Salasnich_2002_eff_wave_eq}, which is the conventional GPE including transverse broadening. Furthermore, we account for the measurement process. The NPSE, its parameters, and the simulation procedure are detailed in the End Matter.\\

\textit{Results $-$}
\begin{figure}
    \centering
    \includegraphics[width=8.6 cm]{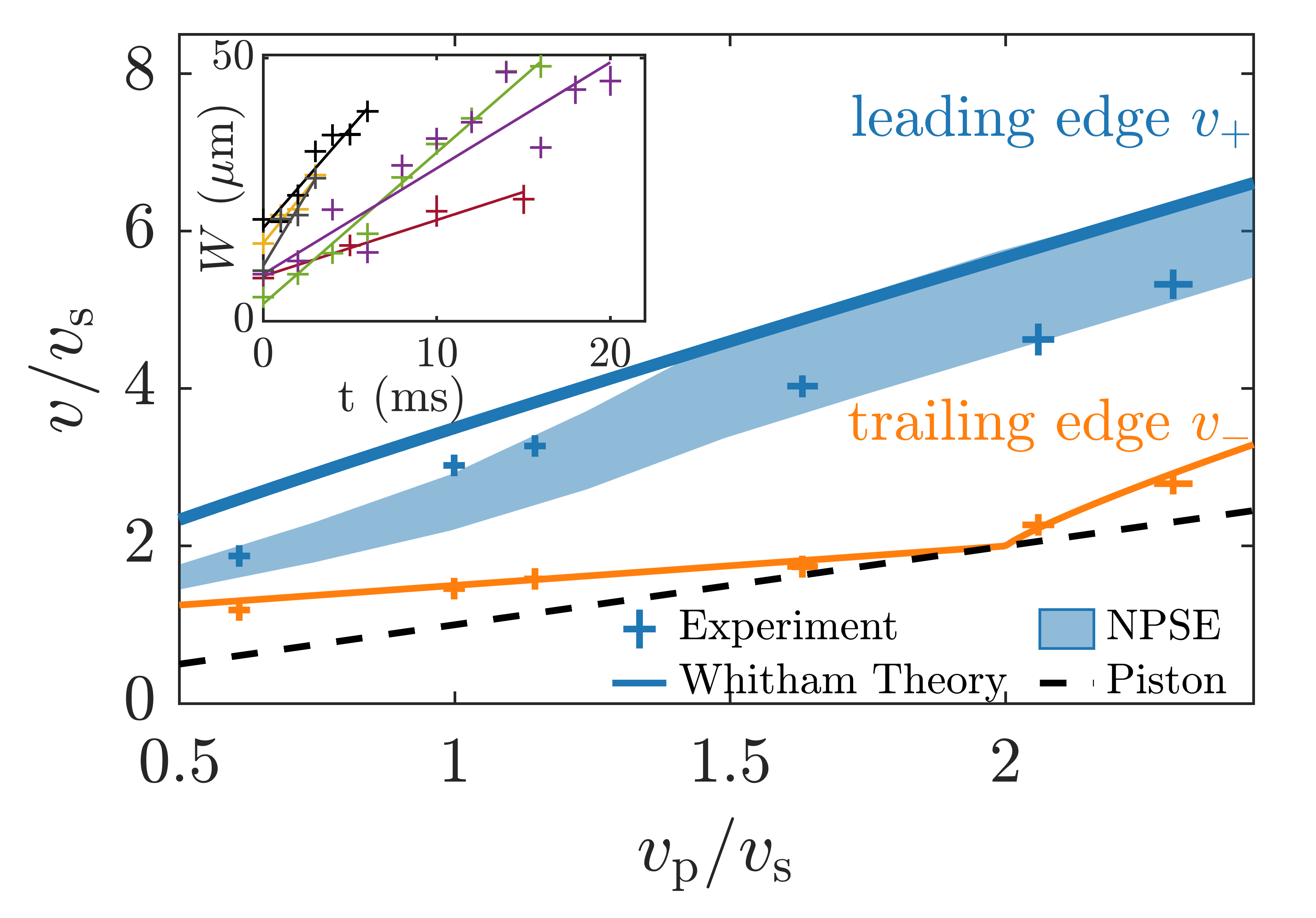}
    \caption{\label{fig:SW_speeds}
    \textbf{Shock region edge speeds for the piston \\protocol.}
    The experimentally measured speed for the leading edge is shown in blue, trailing edge in orange. The corresponding theory values are shown as solid lines in blue and orange respectively. The blue shaded region is obtained via finite temperature NPSE simulations including TOF and imaging. The piston speed $v_\mathrm{p}$ is shown as a dashed black line. Experimental standard errors are derived from statistics of multiple measurements as well as errors extracted from the fitting procedure and are in the range of: $\bar{\sigma}_{v_\mathrm{p}/v_s} \approx0.01-0.07~v_\mathrm{s}$, $\bar{\sigma}_{v_\mathrm{\pm}/v_s} \approx 0.1-0.4~v_\mathrm{s}$.
    \textbf{Inset:} Shock wave widths, $W(t)=z_+(t)-z_-(t)$ and linear fits as solid lines. The widths are colored in order of increasing $v_\mathrm{p}$ as : red, green, purple, yellow, black, gray
    }
\end{figure}
First, we extract and study the propagation velocities of the shock wave fronts across a wide range of piston velocities. The results are presented in Fig.~\ref{fig:SW_speeds}. The measured velocities are consistent with the theoretical predictions of Whitham's method for dispersive shock waves (see End Matter Eq.~\ref{eq:velocity_front_back_DSW}): The velocity of the leading edge $v_+$ asymptotically approaches $2v_\mathrm{p}+2v_\mathrm{s}$. Meanwhile, the trailing edge velocity $v_-$ experiences a kink, a discontinuity in the first derivative, at the critical velocity $v_\mathrm{p}=2v_\mathrm{s}$. Below, it scales as $v_\mathrm{p}/2$ and above $v_-$ approaches a linear growth scaling as $2v_\mathrm{p}$. Crucially, the observed kink and widening shock region are both distinguishing properties of DSWs. The measured leading edge speeds $v_+$ lie systematically below the theoretically predicted values, which we attribute to finite temperature and imaging effects. Comparison with simulations reveals a similar behavior. Error bars are derived from standard errors of the density distribution, as well as errors from the fitting process.\\

\begin{figure}
    \centering
    \includegraphics[width=8.6 cm]{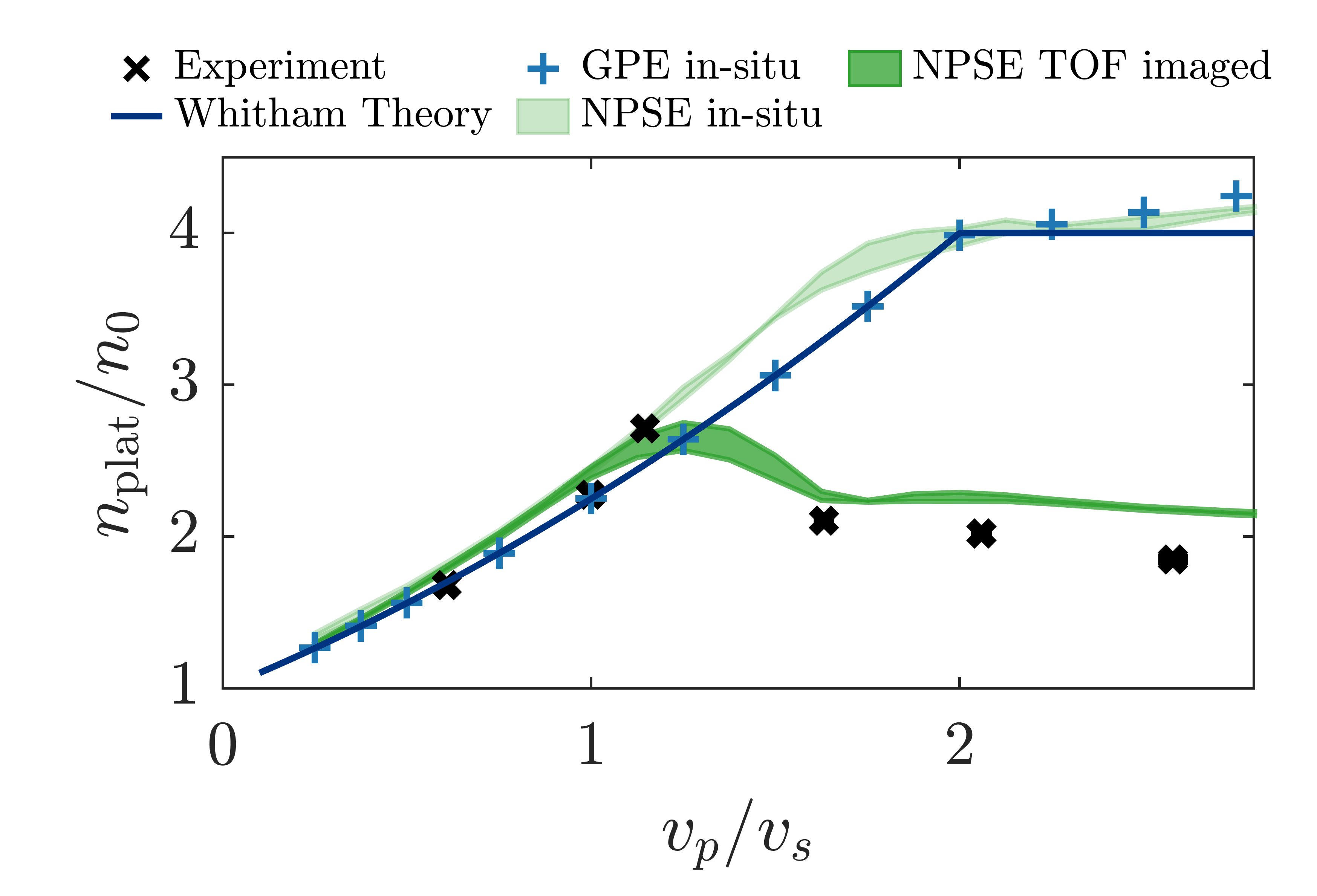}
    \caption{\label{fig:n_max}
    \textbf{Maximum atom accumulations for the piston protocol.}
    Measured plateau densities (black) are compared to Whitham theory predictions (dark blue line) and simulations of increasing complexity: zero temperature GPE in-situ (blue pluses), finite temperature NPSE in-situ (light green region) and finite temperature NPSE simulations including 2 ms TOF and imaging (NPSE TOF imaged - dark green region). Experimental standard errors are in the range $\approx0.02-0.06~n_0$. 
    }
\end{figure}

Next, we study the accumulation of atoms between the piston and the shock region. Whitham's method predicts a region of constant density for $v_\mathrm{p}<2v_\mathrm{s}$, whereby the plateau density $n_{\mathrm{plat}}$ increases quadratically until the critical piston speed $v_\mathrm{p}=2v_\mathrm{s}$, where the piston catches the trailing edge. Beyond, $v_\mathrm{p}=2v_\mathrm{s}$ the plateau is replaced by large density oscillations with a maximum amplitude of $n_{\mathrm{plat}}=4n_0$ independent of $v_\mathrm{p}$ (see End Matter Eq.~(\ref{eq:n_plateau_piston})). The results are shown in Fig.~\ref{fig:n_max}.

For slow piston speeds, $v_\mathrm{p}<1.2v_\mathrm{s}$, the measured $n_\mathrm{plat}$ agrees closely with the predictions of Whitham theory. At larger $v_\mathrm{p}$, however, the observations lie considerably below the predicted values. We assess this discrepancy using simulations of progressively increasing complexity. First, finite-temperature in-situ NPSE simulations account for thermal effects and transverse broadening, isolating deviations arising from the quasi-1D dynamics. The reduced speed of sound leads to a slightly higher plateau density for $v_\mathrm{p}<2v_\mathrm{s}$. We then include the measurement process by accounting for 2 ms TOF expansion and the finite imaging resolution, thereby incorporating phase–density mixing and the coarse-graining of large density oscillations (more details in the Supplementary Material \cite{Supplementary_Material}). This reproduces the observed reduction in maximum density and captures the experimental trend well. At the highest piston speeds, $v_\mathrm{p}>1.6v_\mathrm{s}$, residual deviations in both the edge velocities and maximum densities remain, which we trace to the finite piston height in the next paragraph.\\


\textit{Finite piston $-$} To better understand the experimental dynamics we need to accurately model the implemented piston. Unlike previous works that sweep a laser beam across the condensate \cite{Dissipative_shock_wave_Mossman2018}, we shape the full beam with the DMD and overlap it with the cloud, which reduces the intensity, see Fig.~\ref{fig:exp_piston_problem}(a) (more details in the End Matter). For high piston speeds $v_\mathrm{p}\geq1.6v_\mathrm{s}$, we observe that the local chemical potential starts to exceed the finite dipole potential of the piston, whereby excess atoms are lost from the box. Importantly, these are local spillovers into the larger harmonic confinement, with the atoms still detectable, and thus differ fundamentally from genuine losses in their effect on 1D gases \cite{losses_1d_integr_systems_Bouchoule2020}. We call this piston \textit{leaky}.\\ 
\begin{figure}
    \centering
    \includegraphics[width=8.6 cm]{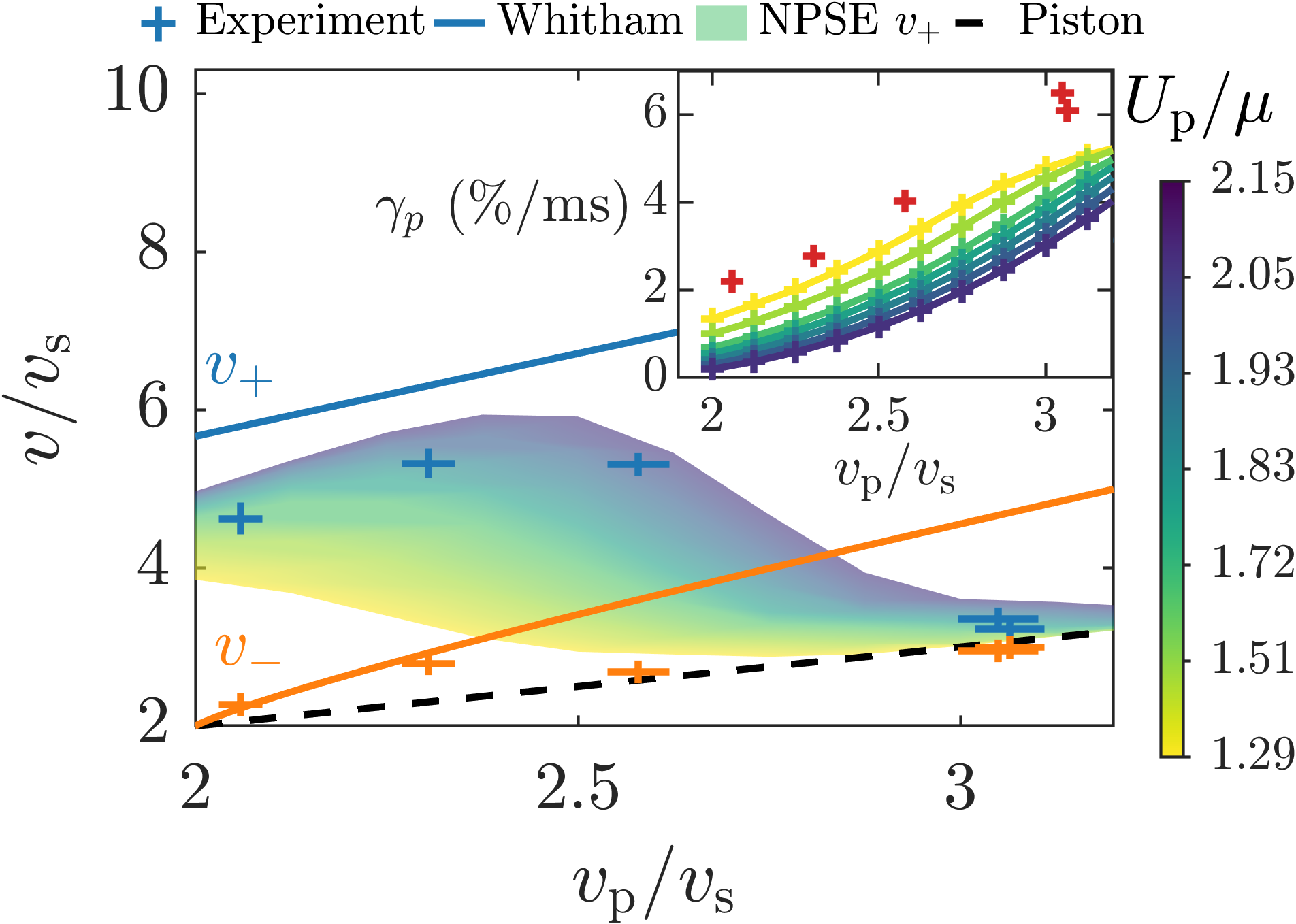}
    \caption{\label{fig:leaky_piston}
    \textbf{Edge velocities for a leaky piston.}
    Experimentally measured leading and trailing edge shock region speeds $v_\pm$ (blue/orange markers) are compared to theory predictions (solid blue/orange lines) and finite temperature NPSE simulations (shaded area $v_+$), including TOF and imaging, with varying piston heights $U_\mathrm{p}$ in relation to the initial chemical potential $\mu$. The shading is intended as a guide to the eye. Standard errors of $v_+/v_\mathrm{s}$ are in the range $0.05-0.1$. 
    \textbf{Inset:} For the same datasets: $\gamma_{p}$: Percent of total atoms that leave the box per time as a function of $v_\mathrm{p}$. Experimental errors are obtained from standard errors of the density distribution as well as errors extracted from the fit. Standard errors of $\gamma_\mathrm{p}$ are in the range $0.05-0.30\%$.
    }
\end{figure}
The measured velocities of the shock wave edges for a leaky piston are presented in Fig.~\ref{fig:leaky_piston}. We compare datasets with the same approximate initial system size $L\approx70 ~\mu\mathrm{m}$ and density $n_{\mathrm{1D}}\approx 70~ \mu \mathrm{m}^{-1}$. These datasets are driven by the same finite piston height (for more details, see the Supplementary Material \cite{Supplementary_Material}), and thus by fixing the ratio of initial chemical potential to $U_\mathrm{p}$ we can observe the dynamics for varying $v_\mathrm{p}$. We find that both shock edge velocities, $v_-$ and $v_+$ exhibit a saturation at $v_\mathrm{p} \approx 2.3 v_\mathrm{s}$. The effect of the finite piston height is most evident at the leading edge of the shock wave, since the highest-momentum particles that would form it instead leave the box. 

To confirm that the observed dynamics are caused by a finite height piston, we compare with NPSE simulations for varying strengths of the piston potential. Here we find a good agreement. The piston loss rate: $\gamma_\mathrm{p}=(1/N)~\mathrm{d}N_\mathrm{p}/\mathrm{d}t$, where $N_\mathrm{p}$ is the number of atoms measured outside of the box, increases approximately linearly with $v_\mathrm{p}$ from $\approx 2.1~\% /$ms to $6~\% /$ms, similar to simulations, albeit systematically higher. We attribute the deviations to systematic imperfections and non-uniformity of the piston potential (for more details, see Supplementary Material \cite{Supplementary_Material}). The genuine atomic loss rates are very low at $\leq0.25 \% $/ms and do not depend on $v_\mathrm{p}$. For very high $v_\mathrm{p} > 3v_\mathrm{s}$, the measured $v_+$ and $v_-$ collapse to the piston speed and we observe a very narrow, almost constant shock region propagating rigidly with the piston, in agreement with simulations (for more details, see the Supplementary Material \cite{Supplementary_Material}). \\

\textit{Transverse effects $-$} For an ideal piston, increasing $v_\mathrm{p}$, raises $\mu \gtrsim\hbar\omega_\perp$, populates transverse modes and eventually leads to topological excitations, producing dissipative-like dynamics as observed in prior piston experiments with lower transverse confinement \cite{Dissipative_shock_wave_Mossman2018}. Notably, we do not observe topological excitations or dissipative dynamics. For $v_\mathrm{p}>2v_\mathrm{s}$ the measured peak densities yield $\mu/\hbar \omega_{\perp} $ in the range of $1.13 -1.39$, while in-situ estimates are higher (simulations predict values in the range of $2.17 - 2.85$), indicating non-negligible transverse effects (see Supplementary Material \cite{Supplementary_Material}). Nevertheless, the longitudinal dynamics remain well-reproduced by quasi-1D NPSE, explicitly including transverse broadening, which underscores the importance of transverse effects \cite{GHD_dim_crossover_Moller_2021, Pauli_blocking_Fede_2022, exp_dim_crossover_ultracold_atoms_Guo_2024}.

In brief, the coarse-grained DSW phenomenology remains observable for three reasons. First, the finite piston height limits the local atom accumulation at high piston speeds, as excess atoms simply exit the box. Second, the over-compressed zone is spatially localized next to the piston and the characteristics that form the leading edge quickly exit this region. Third, the observed time-scales are not long enough for the injected transverse energy, visible as transverse broadening (see Supplementary Material \cite{Supplementary_Material}), to develop into collective 3D dynamics or topological excitations. Especially for large $v_\mathrm{p}$, finite size effects limit the measurement time considerably.\\

\textit{Conclusion $-$}
We presented a systematic experimental study of the piston problem in a 1D Bose gas. Our observations demonstrate the presence of dispersive shock waves for piston speeds up to three times the speed of sound, authenticated by measuring linearly growing shock widths and two distinct shock region speeds that follow the predictions obtained by Whitham's method. Notably, the trailing edge speed displays a kink at the critical piston velocity $v_\mathrm{p}=2v_\mathrm{s}$. Finite-temperature NPSE simulations that include TOF and imaging effects reproduce the measured shock dynamics and peak densities, while behavior at high piston speeds are consistent with a finite-height potential. Remarkably, despite local densities exceeding the strict 1D regime, the coarse-grained DSW phenomenology remains observable, highlighting that the dispersive-shock description survives even as the microscopic dynamics depart from the strictly integrable 1D regime. Our results constitute a controlled, quantitative test of dispersive effects in a quasi-1D quantum fluid.

This letter opens further experimental avenues: Increased optical power would allow to approach the ideal piston, enabling investigation into the expected breakdown of integrability at large piston speeds. Sub-micron optical resolution and interferometric phase readout could reveal the shock microstructure, further progressing the investigation into Whitham's asymptotic predictions and the behavior of dispersive shock waves \cite{quantum_gas_microscopy_review_Gross2021}. Multi-piston protocols enable the study of multiple shock regions and shock-shock interactions \cite{multi_shock_regions_NLS_step_Biondini2006}. The shock wave can be characterized by measuring the local quasi-momentum distribution \cite{local_rap_distr_probing_Dubois_2024}. The introduction of weak disorder combined with entropy measurements can provide insights into prethermalization and relaxation dynamics of nearly integrable quantum systems \cite{prethermalization_Gring2012, prethermalization_Langen2016}.\\

\textit{Acknowledgments $-$} The research at TU Wien was supported by the European Research Council under Horizon Europe (project EmQ, Grant No. 101097858), and by the Austrian Science Fund (FWF) (Grant DOI: 10.55776/P36656) and the Austrian Science Fund (FWF) through the Cluster of Excellence ‘Quantum Science Austria’ (quantA) (Grant DOI: 10.55776/COE1).\\

F.M. and P.S. conceived the idea. P.S. carried out the experiments with contributions by F.M., M.T., F.C., S.-C.J. and N.B.. P.S. performed the data analysis, and simulations. P.S. wrote the paper with contributions from all authors. F.M. and J.S. supervised the project.\\
\newline

\textit{Data availability $-$} The data supporting the findings of this study are openly available in the Zenodo repository [DOI 10.5281/zenodo.21991533] \cite{data_repo_DSW}.
\bibliography{references_2}

%
%

\appendix
   \onecolumngrid \begin{center}\bfseries End Matter\end{center} \twocolumngrid
   
\textit{Technical details of the experimental procedure $-$} We use a sequence of standard cooling and trapping techniques to prepare $N=4-10\times10^3$ atoms at a temperature of $T=20-90$ nK. Temperature is measured via density ripples thermometry, utilizing the mixing of phase and density in TOF, whereby thermal in-situ phase fluctuations translate into density fluctuations (ripples) \cite{density_ripples_Manz_2010}. Crucially, we cool the condensate directly into the box trap to prepare initially thermal states. At $t=0$ ms we start the protocol and after the desired evolution time the condensate expands freely for $2$ ms TOF after which we measure the transverse 2D atomic density distribution via absorption imaging, thereby destroying the system. We utilize an imaging system with magnification $12.39$, pixel size in object space: $1.05~\mu\mathrm{m}$ and an effective numerical aperture of NA$= 0.2$. The estimated point spread function obtained from density ripples thermometry after $12.2$ ms TOF is $\sigma_{\mathrm{PSF}}\approx3~\mu\mathrm{m}$. We obtain $50 - 80$ measurements per evolution time and finally post-select on total atom number, fitting a normal distribution and retaining shots within a $\pm1\sigma$ window ($\pm 0.7\sigma$ for the leaky piston) around the mean and discarding approximately $30\%$ ($50\%$) of shots. Although the atom number determines the local speed of sound and in turn the shock dynamics, we observe no cross-dataset correlation of shock observables and post-selection window.\\

\textit{Creating arbitrary longitudinal potentials $-$}
More details regarding the creation of arbitrary longitudinal potentials are found in Ref.~\cite{Tajik:19}. Here we summarize the setup for a better understanding of the resulting finite piston power. We utilize a $\lambda_{\mathrm{dip}}=766.8~\mathrm{nm}$ blue detuned (with respect to the D2 transition of $^{87}$Rb at $\lambda_{\mathrm{D2}}=780.24~\mathrm{nm}$) Gaussian beam whose horizontal extension is elongated by a cylindrical telescope to illuminate the DMD, a SuperSpeed V-9501 Module from ViALUX, featuring 1080 by 1920 micromirrors each sized $10.8~\mu\mathrm{m}$. The light which is reflected by the DMD in the "ON" position is superimposed on the transverse imaging path of the experimental setup via a polarizing beam splitter cube. The horizontal x-axis of the DMD matches the longitudinal z-axis of the condensate. Two demagnification stages, including a lens on a motorized stage, are used to image and focus the DMD pattern in the plane of the atoms, resulting in a final pixel size in the plane of the atoms of $0.45~\mu\mathrm{m}$. As the transverse harmonic oscillator ground state width in the trap is $l_{\mathrm{ho}}\approx0.3~\mu\mathrm{m}$, smaller than the DMD pixel size, we use a horizontal slit in the Fourier plane of the first demagnification stage to implement grayscaling via spatial filtering, which reduces the total intensity. We optimize DMD patterns via iteratively comparing the in-situ atomic density to the desired potential and updating pixel states until convergence. To offset the weak longitudinal harmonic confinement, the box pattern displays the inverse profile with high pixel density at the center decreasing towards the edges of the box. The walls are implemented using the maximally available intensity, that is columns with all pixels on. The piston is then implemented by successively switching on columns of pixels atop this box, starting at the wall, resulting in a reduced effective power towards the center of the condensate, where many pixels are already active. Overall, the piston power is reduced by grayscaling and exhibits longitudinal non-uniformity due to the elongated Gaussian beam and the shared use of DMD for both box and piston.\\

\textit{Theoretical description (NPSE) $-$} The dynamics of a quasi-1D gas of $N$ bosons with repulsive contact interactions is well described by the non-polynomial Schrödinger equation (NPSE) \cite{Salasnich_2002_eff_wave_eq}, which differs from the conventional Gross-Pitaevskii equation (GPE) \cite{pitaevskii2016bose} by including transverse Gaussian broadening of the wave-function: 
\begin{equation}
i\hbar\partial_t\Psi = \left[-\frac{\hbar^2}{2m}\partial_z^2+V+\hbar \omega_{\perp}\sqrt{1+2a_\mathrm{s}|\Psi|^2}\right]\Psi  \text{.}
    \label{eq:NPSE}
\end{equation} 
where $\Psi(z,t)$ is the order parameter, $V(z,t)$ the external longitudinal potential (we leave out the $(z,t)$-dependence for readability), $\hbar$ the reduced Planck constant, $m$ the atomic mass, $\omega_{\perp}$ the transverse trapping frequency and $a_\mathrm{s}$ the scattering length. Writing the GPE in the Madelung representation (More Details see Supplementary Material \cite{Supplementary_Material}) reveals a term $\propto \partial_z^2 |\Psi|$, which is often called \textit{quantum pressure} and becomes relevant on length scales smaller than the healing length $\xi_\mathrm{h}=\hbar/\sqrt{2mn_{\text{1D}}g_{\text{1D}}}$, regularizing the solution at the shock. The effective 1D interaction strength is denoted by $g_{\mathrm{1D}}=2\hbar \omega_\perp a_\mathrm{s} /\sqrt{1+2a_\mathrm{s} n_{\mathrm{1D}}}$. The homogeneous 1D (background) density $n_{\text{1D}}$ determines the speed of sound: 
\begin{equation}
v_\mathrm{s}=\sqrt{\frac{\hbar \omega_{\perp}a_\mathrm{s} n_{\text{1D}}(2+3a_\mathrm{s}n_{\text{1D}})}{m(1+2a_\mathrm{s}n_{\text{1D}})^{3/2}}} \;.
\label{eq:speed_of_sound}
\end{equation}
Note that both $g_\mathrm{1D}$ and $v_\mathrm{s}$ are slightly lower compared to the conventional GPE definitions due to the effects of the transverse broadening.\\

\textit{Whitham DSW observables $-$}
Whitham's Modulation Theory applied to zero temperature GPE for the piston protocol results in a growing shock region where the shock edge velocities can be found by enforcing boundary conditions with the dispersionless region \cite{nonlinear_DSW_whitham_1965, linear_and_nonlinear_waves_whitham2011_BOOK, Piston_DSW_analytics_Hoefer_2008, dispersive_shock_waves_mod_theory_el_hoefer_2016}:
\begin{align}
v_+ &=\frac{2v_\mathrm{p}^2+4v_\mathrm{p}v_\mathrm{s}+v_\mathrm{s}^2}{v_\mathrm{p}+v_\mathrm{s}} \label{eq:velocity_front_back_DSW} \; ,\\
v_- &=0.5v_\mathrm{p}+v_\mathrm{s} ~~~~~~~~~~~~~~~~~~~~~~~~~~~~~~~~~\text{for}~v_\mathrm{p}<2v_\mathrm{s} \nonumber \; \\
v_- &= v_\mathrm{p}+\frac{2v_\mathrm{s}(v_\mathrm{p}-2v_\mathrm{s})K(M)}{v_\mathrm{p}E(M)-(v_\mathrm{p}-2v_\mathrm{s})K(M)}~~~\text{for}~v_\mathrm{p}\geq2v_\mathrm{s}  \nonumber \; ,
\end{align}
where \textit{K} and \textit{E} are the complete elliptic integrals of first and second kind respectively, of $M=4v_\mathrm{s}^2/v_\mathrm{p}^2$. 
The plateau density follows
\begin{align}
\label{eq:n_plateau_piston}
n_{\text{plat}}&=\left(\frac{v_\mathrm{p}}{2}\sqrt{\frac{m}{g_{\text{1D}}}}+\sqrt{n_0} \right)^2  ~~~~~~~~\text{for}~v_\mathrm{p}<2v_\mathrm{s}  \; \\
n_{\text{plat}}&=4n_0~~~~~~~~~~~~~~~~~~~~~~~~~~~~~~~\text{for}~v_\mathrm{p}\geq2v_\mathrm{s}  \nonumber \; .
\end{align}\\

\textit{Numerical simulations $-$} 
We use a standard Fourier-split-step method to simulate the NPSE and GPE dynamics.
First, ground states are obtained via imaginary time propagation and used as initial guesses for further stochastic GPE/NPSE: We evolve the conventional GPE/NPSE with an additional noise term that simulates the system being in contact with a thermal bath, whose correlations are related to the final temperature, arriving at thermal initial states. The box potential is modeled via two finite-sized error function walls on both sides. The piston height is chosen at five times the chemical potential to exclude atoms leaking over the piston, except in the cases explicitly testing the effects of the finite piston height. The piston is implemented by moving one of the walls at constant speed starting at $t=0$. Finally we obtain statistics from 100 realizations. In order to account for TOF and the imaging process we allow for a free propagation and afterwards convolve with a Gaussian, simulating a point spread function of $\sigma_{\mathrm{PSF}}=3 ~\mu\mathrm{m}$ consistent with our density ripples measurements. Notably, our simulations do not account for systematic box imperfections, experimental drifts of atom number and temperature, or for longitudinally varying piston power.

\end{document}